\documentclass[sigconf,nonacm]{acmart}
\setcopyright{none}
\renewcommand\footnotetextcopyrightpermission[1]{}
\AtBeginDocument{%
  }

\usepackage{graphicx}   %
\usepackage{booktabs}   %
\usepackage{multirow}   
\usepackage{acro}
\usepackage{hyperref}
\usepackage{balance}

\DeclareAcronym{LLM}{
  short = LLM ,
  long  = large language model
}

\newcommand{\inlinequote}[1]{``\textit{#1}''}

\begin{document}

\title{When Should Teachers Control AI Generation for Mathematics Visuals?}

\author{Zhengxu Li}
\orcid{0009-0002-0620-4090}
\authornote{These authors contributed equally to this research.}
\affiliation{%
  \institution{ETH Zurich}
  \city{Zurich}
  \country{Switzerland}
}
\email{zhengxli@ethz.ch}

\author{Junling Wang}
\orcid{0000-0002-4526-2907}
\authornotemark[1]
\affiliation{%
  \institution{ETH Zurich}
  \city{Zurich}
  \country{Switzerland}
}
\email{junling.wang@ai.ethz.ch}

\author{April Yi Wang}
\orcid{0000-0001-8724-4662}
\affiliation{%
  \institution{ETH Zurich}
  \city{Zurich}
  \country{Switzerland}
}
\email{april.wang@inf.ethz.ch}

\renewcommand{\shortauthors}{Zhengxu Li, Junling Wang, and April Yi Wang}

\begin{abstract}
Generative AI has the potential to help teachers rapidly create classroom-ready visual materials, particularly in mathematics where diagrams and visual representations must be pedagogically meaningful and instructionally correct.
However, current generative tools primarily support prompting and post-hoc editing, leaving open a key question for correctness-sensitive educational authoring: when in the generation pipeline should teachers exert control?
In this paper, we investigate how the timing of human control in AI-assisted generation shapes teachers’ visual authoring practices in correctness-sensitive tasks. 
We introduce a design space of three stages of control: \emph{pre-generation control}, where users specify intent solely through natural language prompts before generation; \emph{mid-generation control}, where users inspect and confirm an explicit layout structure before the system completes generation; and \emph{post-generation control}, where users directly modify AI-generated visuals after generation through object-level edits. 
In a within-subject, mixed-methods study with 24 primary mathematics teachers, post-generation control received higher ratings on predictability and correctness, while other subjective measures showed no reliable differences. 
Qualitative findings explain these differences by revealing workflow trade-offs: highly automated, pre-generation control supports rapid ideation but reduces perceived agency and predictability; mid-generation control improves structural alignment at the cost of additional effort; and post-generation control preserves user agency through low-cost, direct verification and correction. 
Together, these results suggest that in correctness-sensitive educational tasks, effective generative tools should align system behavior with teacher intent and support stage-dependent workflows that combine automation with direct manipulation.
\end{abstract}


\begin{teaserfigure}
  \includegraphics[width=\textwidth,trim={1cm 0.5cm 1cm 1cm},
  clip]{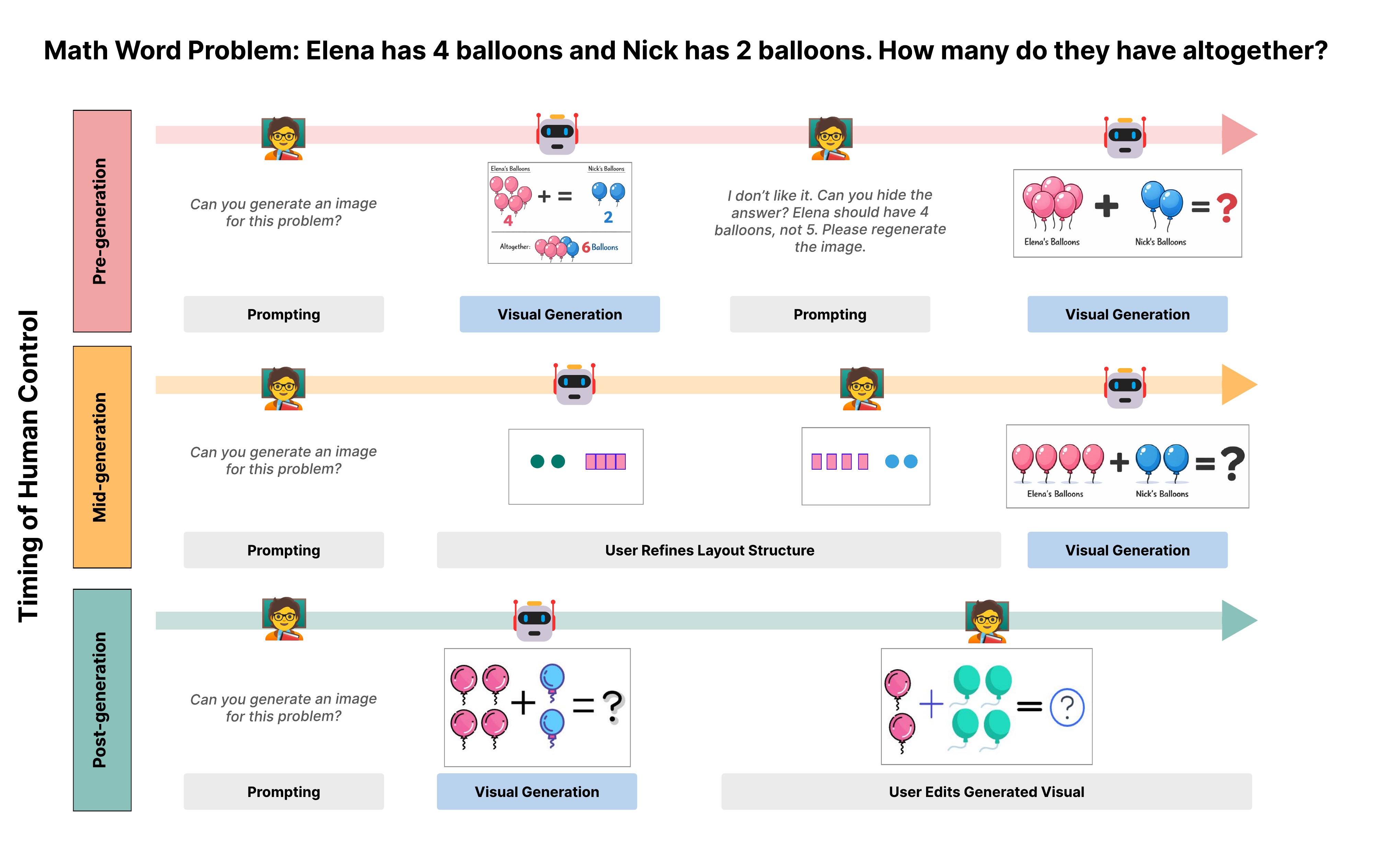}
  \caption{Abstract workflows of the three implemented interaction paradigms, namely pre-generation control, mid-generation control, and post-generation control, illustrated using the same math word problem (“Elena has 4 balloons and Nick has 2 balloons. How many do they have altogether?”). The paradigms differ in when teacher control occurs relative to AI output.}
  \Description{workflow diagram comparing three AI-assisted visual authoring paradigms: pre-generation prompting, mid-generation layout confirmation, and post-generation object-level editing.}
  \label{fig:teaser}
\end{teaserfigure}


\maketitle

\section{Introduction}

Visual representations play a central role in primary mathematics education, helping students reason about quantities, relationships, and operations that are difficult to express through text alone \cite{duval1999representation,rivera2011toward,haylock2007key}. Well-designed visuals can improve comprehension and problem-solving, especially for novice learners \cite{rau2017conditions,bobek2016creating,wu2004exploring}. Yet creating pedagogically aligned visuals requires time, design effort, and appropriate tools \cite{xu2021procedural}. In practice, many teachers rely on static textbook images or generic online resources that may not fit specific lesson goals or classroom contexts \cite{clark2010graphics,horton2003learning,abdallah2011web}---a constraint that becomes more salient in Learning at Scale contexts, where instructional materials must be generated, adapted, and reused across diverse classrooms and large learner populations \cite{holmes2019artificial,koedinger2015learning}. In such settings, even small inaccuracies can propagate widely, amplifying the need for workflows that are both scalable and reliable.

Recent advances in generative AI, especially text-to-image models, offer new possibilities for the creation of customized instructional visuals \cite{ramesh2021zero,rombach2022high,midjourney}. Prior research has explored prompting strategies and multimodal interaction techniques to help educators steer such systems \cite{ali2024picture,liu2022design}. However, classroom use remains challenging: effective prompting requires expertise, system behavior can be opaque, and verifying alignment with pedagogical intent is often difficult \cite{long2020ai,zamfirescu2023johnny,holmes2021ethics,mogavi2024chatgpt}. These issues are amplified in mathematics: representations must faithfully reflect quantities and relations, yet current models frequently fail at counts and structured constraints \cite{zhang2024mathverse,hou2024vision,choudhury-etal-2025-vision}. As a result, fully automated generation is ill-suited for scalable deployment without low-cost mechanisms for verification and correction. While similar issues have been studied in the text generation area \cite {bommasani2021opportunities,wei2022chain}, less is known about how they manifest in visual generation workflows, particularly in correctness-sensitive domains such as mathematics education.

These limitations highlight a broader design challenge: unlike traditional authoring software, generative tools create a multi-stage human--AI workflow. Generation is decomposed into steps such as intent specification, intermediate planning, and localized refinement \cite{lin2023layoutprompter,chung2023promptpaint}. This makes \textbf{when teachers can intervene} a central interaction design question. Human control may occur \textit{before} generation through intent specification, \textit{during} generation through inspection or editing of intermediate structure, or \textit{after} generation through direct correction of outputs. While emerging tools support parts of this design space \cite{wang2024promptcharm,almeda2024prompting}, we lack empirical understanding of how \textbf{the timing of human control} shapes educators' workflows, perceived agency, and ability to ensure instructional correctness. To address this gap, we ask the following research questions:

\textbf{RQ1}: How does the timing of human control affect the correctness of AI-generated instructional visuals?

\textbf{RQ2}: How does control timing influence teachers’ perceived predictability, trust, and sense of agency in the generation process?

\textbf{RQ3}: What trade-offs emerge between automation, effort, and workflow efficiency across different control timings?

To answer these questions, we present a web-based research prototype\footnote{Supplementary materials: \url{https://osf.io/c4ebf}; Web prototype: \url{https://github.com/ViktorLee426/Visual4Math}} to operationalize different control timings in AI-assisted visual authoring for primary mathematics problems. We compare three interaction paradigms: (1)\emph{pre-generation control}, where teachers specify intent through conversational prompting before visual is generated; (2)\emph{mid-generation control}, where teachers inspect and confirm an explicit layout structure before the system completes generation; and (3)\emph{post-generation control}, where teachers directly modify AI-generated visuals through object-level manipulation(Figure~\ref{fig:teaser}). We conducted a within-subject, mixed-methods study with \textbf{24} primary mathematics teachers, who used all three paradigms to create visuals for standardized arithmetic problems. We analyzed teachers' workflows, perceived agency, and the classroom readiness of generated visuals. Our contributions are:
\begin{itemize}
    \item We introduce \textbf{control timing} as a key design dimension in AI-assisted visual authoring, and operationalize it through three different interaction paradigms.
    
    \item Through a mixed-methods study with 24 teachers, we provide empirical evidence showing how control timing shapes trade-offs between correctness, predictability, and effort.
    
    \item We derive design implications for generative educational tools, and connect these findings to broader human--AI interaction principles for scalable, reliable content creation.
\end{itemize}


\section{Related Work}

\subsection{Visual Authoring in Education}
A substantial body of research in the learning sciences shows that visuals can improve learning outcomes when they are aligned with instructional intent \cite{ardasheva2018representation,guo2020impact,renkl2017studying}. Prior work explains these benefits through dual coding and multimedia learning, where complementary verbal and visual representations support mental model construction under limited cognitive resources \cite{clark1991dual,mayer2005cognitive}. In mathematics, representation choices shape how learners interpret quantities and relationships, with work on \emph{concreteness fading} highlighting transitions from concrete to abstract representations \cite{fyfe2014concreteness}.

Crucially, instructional visuals must be correct and interpretable. Visuals that misrepresent quantities or relations can introduce misconceptions rather than support learning \cite{cox1996analytical,macdonald1977numbers}. Prior work also suggests that the construction of representations can itself be educational, as it improves learning outcomes in several contexts \cite{bobek2016creating}. Together, these findings highlight a key requirement for teacher-facing tools: supporting visuals that are structurally faithful to underlying mathematical relationships \cite{rich2020teacher,miller2021motivating}.

While this literature establishes why visuals matter for learning, it provides limited insight into how teachers can efficiently create appropriate visuals under real classroom constraints, or how contemporary generative systems should be integrated into the workflow \cite{pratschke2024generative,aperstein2025generative}. Our work addresses this gap by studying teacher-facing visual creation for primary mathematics problems and by examining designs that support correctness-oriented authoring.

\subsection{Text-to-Image Generation}

Text-to-image models enable rapid generation from natural language prompts and have substantially lowered the barrier to visual authoring \cite{ko2023large}. Research on prompt engineering shows that output quality is highly sensitive to linguistic choices and provides actionable guidance for improving results through careful prompt construction \cite{liu2022design,khan2024quick,geroimenko2025essential}. In educational contexts, these capabilities suggest the potential for customizable visuals that teachers can adapt to specific lesson objectives or learner needs \cite{almuhanna2025teachers}.

At the same time, educational applications, particularly in mathematics, expose fundamental limitations of prompt-only generation. Text-to-image models often struggle with object counts and spatial relationships \cite{ray2023cola,wu2024conceptmix}. Recent evaluations show that performance degrades sharply when prompts require precise quantities or compositional constraints \cite{kajic2024evaluating,you2024depicting}. These limitations have motivated a growing shift from prompt-based, one-shot generation toward multi-stage generation pipelines that include stages such as intent specification, intermediate planning, and iterative refinement. This creates new opportunities for interaction design as users can intervene at multiple stages of the generation process \cite{almeda2024prompting}. 


A parallel line of work has examined human control and verification in text generation systems, where issues such as hallucination, controllability, and user trust have been widely studied \cite{bommasani2021opportunities,shneiderman2022human}. However, visual generation introduces distinct challenges: errors are harder to localize, correctness depends on spatial and quantitative relationships, and intermediate reasoning is less transparent \cite{hussen2025advancing}. As a result, interaction paradigms developed for text generation do not directly transfer to visual authoring, motivating the need to study control mechanisms tailored to visuals.

\subsection{Interaction Paradigms for Visual Creation}

Building on these technical advances, researchers have begun exploring how different interaction paradigms shape users' experiences with generative visual systems. Rather than treating text-to-image generation as a one-shot process, recent work emphasizes workflows that support intent expression, failure diagnosis, and refinement. Two recurring lines of research are particularly relevant.

\textbf{Augmenting prompting with richer inputs and refinement loops.}
Several systems extend prompting beyond raw text by supporting multimodal intent specification and iterative refinement. For example, DesignPrompt enables moodboard-style prompt composition to support design exploration \cite{peng2024designprompt}, while PromptCharm provides iterative prompt refinement and image editing mechanisms that reduce trial-and-error for novice users \cite{wang2024promptcharm}. Other systems emphasize structured exploration and comparison across many generations: DreamSheets uses spreadsheet-style operations to organize and analyze prompt variations \cite{almeda2024prompting}, and Luminate supports systematic exploration of the design space to avoid premature convergence \cite{suh2024luminate}. Collectively, these systems illustrate how prompt-centered iteration and comparison can shape users’ sense of control and exploratory behavior.

\textbf{Imposing structure and spatial control.}
A complementary line of work introduces explicit structure, such as layouts or region-based constraints, to improve compositional control. Examples include LayoutPrompter \cite{lin2023layoutprompter} and SketchFlex \cite{lin2025sketchflex}, as well as related work in 3D layout generation \cite{wang2025chat2layout,tam2024scenemotifcoder}. While effective in creative domains, educational visual authoring introduces additional requirements such as numerical correctness and interpretability \cite{tian2024enhancing}.

Across these interaction paradigms, systems differ in both the mechanisms and the \emph{timing} of user intervention. However, little empirical work compares how the timing of control shapes teachers’ ability to ensure correctness within real classroom workflows. We address this gap by examining how different control timings influence educators’ capacity to verify, correct, and  deploy generated mathematics visuals.

\section{Experimental System}

To investigate how the timing of human control influences teachers’ visual authoring workflows, we developed a web-based research prototype that instantiates three interaction conditions within the same domain. All conditions support the creation of visuals for primary mathematics word problems, while differing systematically in when teachers intervene during the generation process. By holding instructional goals and problem content constant, the system enables a comparison of three control timings: \emph{Pre-generation control}, \emph{Mid-generation control}, and \emph{Post-generation control}.



\subsection{Design Rationale for Interaction Conditions}

\subsubsection{Pre-generation Control}
\emph{Pre-gen} represents prompt-first workflows in which teachers specify intent in natural language before each generation step. We include this condition because it reflects how widely-deployed text-to-image tools are used in practice and allows us to observe how educators translate correctness-sensitive goals into prompts, as well as how ambiguity in language affects perceived reliability and classroom readiness\cite{liu2022design,zamfirescu2023johnny,wang2024promptcharm}.


\subsubsection{Mid-generation Control}
\emph{Mid-gen} represents workflows that expose an intermediate structure that teachers can inspect and adjust before final rendering. This condition tests whether providing an explicit, editable intermediate representation helps teachers express and verify structural intent (e.g., grouping and spatial relations) earlier in the pipeline, potentially reducing downstream correction and increasing confidence \cite{lin2023layoutprompter,wang2025chat2layout,shneiderman2022human}.


\subsubsection{Post-generation Control}
\emph{Post-gen} represents edit-after workflows in which AI provides an initial draft and teachers retain full control through direct manipulation to reach a correct final visual. We include this condition to examine the trade-off between transparency and effort in correctness-sensitive authoring: post-hoc edits can support precise verification and correction, but may increase manual work compared to earlier-stage control \cite{shneiderman1983direct,norman1986cognitive,wang2025generating}.


\subsection{Workflow Example}

\begin{figure*}[t]
\centering
\includegraphics[width=\textwidth]{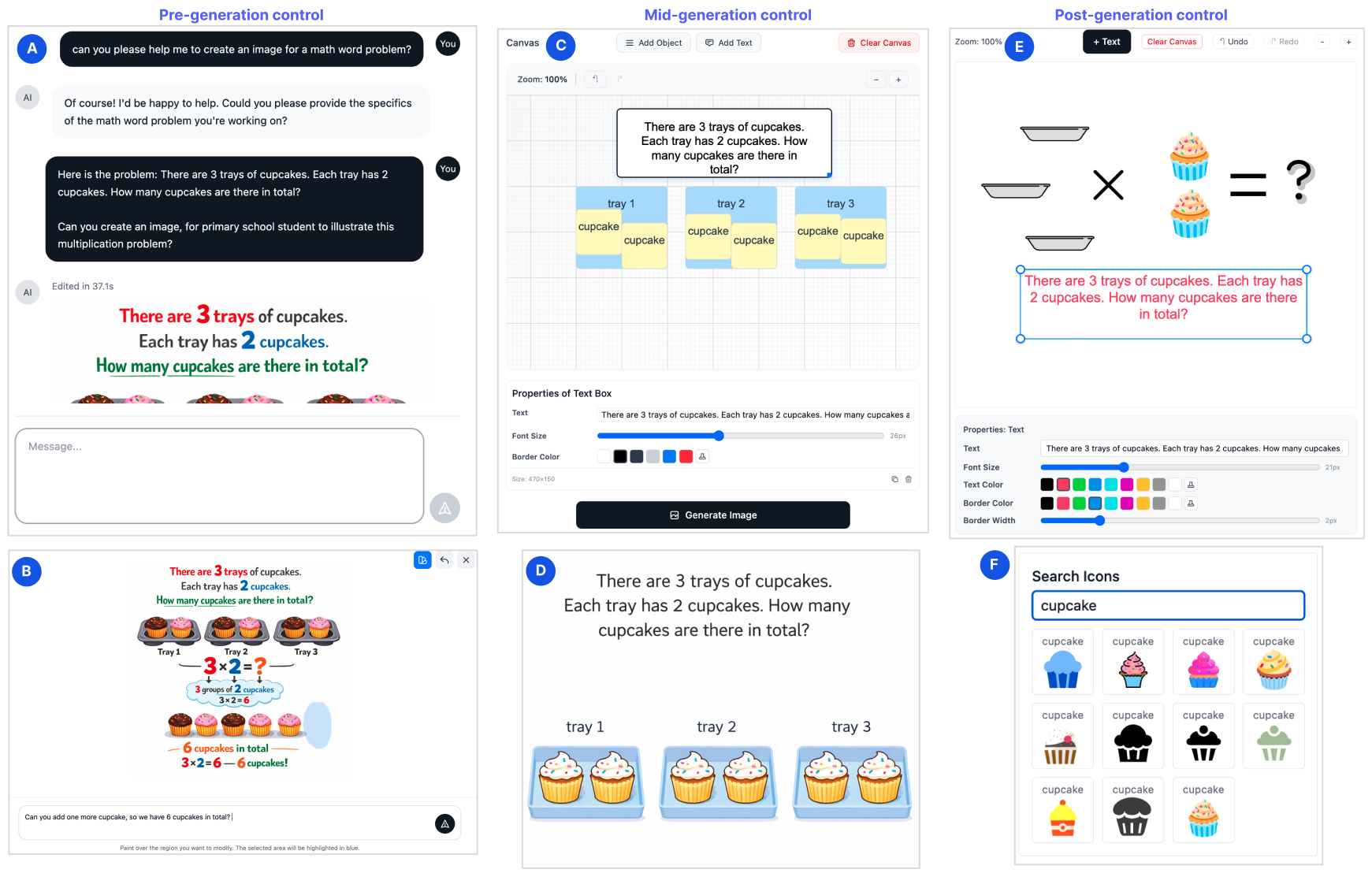}
\caption{
Workflow comparison across three interaction paradigms using the same multiplication problem.
Pre-gen(A,B): conversational prompting and region-level correction.
Mid-gen(C,D): editable layout followed by AI rendering.
Post-gen(E,F): AI-initialized draft followed by object-level direct manipulation.
}
\label{fig:workflow}
\end{figure*}

To illustrate how the timing of control shapes the authoring process, we consider the same primary mathematics word problem across all three tools: \emph{There are 3 trays of cupcakes. Each tray has 2 cupcakes. How many cupcakes are there in total?} The task and problem content remain identical; only the interaction paradigm differs (Figure~\ref{fig:workflow}).

The teacher begins by interacting with \textbf{Pre-gen}, which has a conversational interface similar to modern \ac{LLM} web interfaces (Figure.~\ref{fig:workflow}A). The teacher first asks the system to help create an image for a math word problem. After the system requests clarification, the teacher provides the full problem statement and asks for a visual suitable for primary school students. The system generates an image directly within the chat history. Upon inspecting the output, the teacher notices that the generated image contains an incorrect number of cupcakes. As shown in Figure.~\ref{fig:workflow}B, the teacher selects a region of the image using a brushing tool and provides a localized correction (``add one more cupcake so we have 6 in total''). The system regenerates the image accordingly. Importantly, in this paradigm, control occurs exclusively before each generation step: once a prompt or edit instruction is submitted, the internal generation process is opaque to the teacher. There is no access to intermediate structure or object-level manipulation.

The teacher then tries \textbf{Mid-gen} and again enters the same math word problem. Instead of immediately producing a final image, the system first generates an editable layout (Figure~\ref{fig:workflow}C). The layout consists of labeled bounding boxes representing visual elements (e.g., trays and cupcakes) and text boxes for the problem statement. In this example, the system proposes three tray containers, each containing two cupcakes. The teacher can adjust object and text boxes—including positions, quantities, sizes, and colors—before image synthesis. Once satisfied with the layout, the teacher submits it for rendering. The system then generates a final image conditioned jointly on the original math problem and the edited layout specification (Figure~\ref{fig:workflow}D). Here, human control occurs during generation: the teacher intervenes at the intermediate structural stage that constrains the final output.

Finally, the teacher uses \textbf{Post-gen}. The system generates an initial editable visual configuration composed of SVG elements placed directly on a canvas (Figure~\ref{fig:workflow}E). Each tray and cupcake is represented as an individual SVG object. The teacher can directly resize, duplicate, delete, or reposition elements, add text boxes, and modify the font and color of text. If different visual styles are desired, the teacher can search an integrated icon library (Figure.~\ref{fig:workflow}F) and drag new SVG elements onto the canvas. Unlike Pre-gen and Mid-gen, no further automated image synthesis occurs after AI initialization. All subsequent changes are deterministic and immediately visible. This paradigm affords full object-level control, enabling incremental correction and alignment with pedagogical intent.

\subsection{Technical Implementation}

Our three tools leverage multimodal large language models (MLLMs) to implement different interaction conditions for educational visual generation. We selected state-of-the-art models available at implementation time: GPT-5 for text understanding and intent analysis, and GPT-image-1.5 for image generation. 

\textbf{Pre-gen} provides a conversational interface that routes user inputs based on detected intent categories. An MLLM classifies requests into (1) brainstorming, (2) image generation, or (3) image modification. Depending on the category, the system forwards the user inputs( with optional image or mask regions), to the appropriate text or image API endpoint. Conversation history, including previously generated images, is maintained in the backend and passed with each API call to support coherent multi-turn interaction.

\textbf{Mid-gen} introduces an intermediate layout stage before image synthesis. The math word problem is first processed by LLM to produce a structured layout in JSON format, representing objects as rectangular boxes. The layout is rendered on an interactive canvas where users can adjust positions, sizes, colors, and labels. Upon confirmation, the system converts the edited layout into a text specification and screenshot, which are provided together with the original problem text to GPT-image-1.5 for final image generation.

\textbf{Post-gen} adopts an edit-after workflow. We use \textsc{Math2Visual}~\cite{wang2025generating} to convert math word problems into structured visual specifications (JSON), which are deterministically rendered as editable elements on a web-based canvas. Teachers directly manipulate these elements without further model intervention. Minor interface adaptations were applied while keeping the underlying pipeline unchanged.

The system is implemented as a web application. A FastAPI backend orchestrates model calls and stores session data, logs, and generated images in SQLite. The frontend (React) communicates with the backend via RESTful APIs to support real-time interaction.

\section{User Study}

We conducted a within-subject, mixed-methods study with 24 primary mathematics teachers to examine how three AI-assisted visual authoring paradigms shape teachers' workflows, sense of control, and perceptions of correctness. 
Participants used all three paradigms to create visuals for arithmetic word problems in remote think-aloud sessions. We collected questionnaires, interaction logs, generated images, and semi-structured interviews for analysis.

\subsection{Participants and Recruitment}

We conducted a prior power analysis using G*Power\cite{faul2009statistical} for a repeated-measures design with three within-subject conditions. Assuming an effect size of $f=0.6$, a significance level of $\alpha = 0.05$, power of $0.8$, the analysis indicated a required sample size of 24 participants. 
We recruited primary-level mathematics teachers (9 male, 15 female; mean age = 37.5 years; mean teaching experience = 10.54 years) via Prolific, requiring fluent spoken English.
Participants taught across Grades 1--6, with all teachers having experience teaching early primary levels (Grades 1--2) and arithmetic word problems, ensuring familiarity with the type of instructional scenarios used in this study. All had at least two years of teaching experience and regularly used images for instruction (weekly to daily). Additional demographic information is reported in Table~\ref{tab:participant_background}.

\begin{table*}[t]
\centering

\caption{Participant demographics, including participant ID (PID), country, teaching experience(years), frequency of AI use, duration of AI experience and image creation tools previously used.}

\label{tab:participant_background}
\scriptsize
\resizebox{\textwidth}{!}{%
\begin{tabular}{c c c l l p{5cm}}
\hline
\textbf{PID} & \textbf{Country} & \textbf{Teaching experience} & \textbf{AI use frequency} & \textbf{AI experience}& \textbf{Image creation tools used} \\
\hline
1  & UK & 6 & daily or multiple times a day & 1--2 years & Figma; Powerpoint; Canva; Midjourney; ChatGPT \\
2  & US & 8 & a few times a month & 6 months--1 year  & Powerpoint; Canva; ChatGPT \\
3  & US & 10 & a few times a week & more than 3 years  & Google Slides; Powerpoint; ChatGPT \\
4  & UK & 25 & almost daily & 1--2 years  & Powerpoint; Canva; ChatGPT \\
5  & US & 27 & daily or multiple times a day & 6 months--1 year & Google Slides; ChatGPT \\
6  & UK & 2 & daily or multiple times a day & more than 3 years &  Google Slides; Powerpoint; ChatGPT \\
7  & South Africa & 5 & a few times a week & 1--2 years & Adobe Illustrator; Google Slides; Canva \\
8  & US & 15 & daily or multiple times a day & 1--2 years & Google Slides; Gemini \\
9  & Australia & 15 &  about once a week & 6 months--1 year & Powerpoint; Canva \\
10 & UK & 18 & a few times a week & 1--2 years &  Powerpoint; ChatGPT \\
11 & US & 21 & a few times a week & 1--2 years &  Google Slides; Canva; ChatGPT \\
12 & India & 14 & almost daily & 6 months--1 year & Sketch; Powerpoint \\
13 & Canada & 7 & a few times a month & 1--2 years & Google Slides; Canva; ChatGPT \\
14 & UK & 15 &a few times a month & 1--2 years &  Google Slides; Canva; ChatGPT \\
15 & South Africa & 4 &a few times a week & 1--2 years &  Paint; Google search images \\
16 & Kenya & 3 & almost daily & 1--2 years &  Google Slides; Powerpoint \\
17 & UK & 3 &a few times a month & less than 6 months &  Powerpoint; Canva \\
18 & Mexico & 3 &about once a week & 2--3 years &  Sketch; Google Slides; Powerpoint; Canva; ChatGPT \\
19 & UK & 7& about once a week & 1--2 years & Powerpoint; Canva \\
20 & France & 12 & a few times a week & 1--2 years  & Adobe Photoshop; Powerpoint; Canva; Gemini \\
21 & UK  &  26 & about once a week & 1--2 years  & Google Slides; Powerpoint \\
22 & UK & 6 & a few times a week & 6 months--1 year & Adobe Photoshop; Powerpoint \\
23 & Greece & 6  & a few times a month & 1--2 years & Powerpoint; Canva \\
24 & Portugal & 8 & less than once a month & 1--2 years & Adobe Illustrator; Powerpoint \\
\hline
\end{tabular}%
}
\end{table*}

\subsection{Study Protocol}
All study sessions were conducted remotely via Zoom and lasted approximately 70 minutes. Participants received \pounds14 as compensation. During the session, participants shared their screens and were instructed to think aloud while completing the tasks.

At the start of the session, participants received a brief overview of the study procedure. They then used all three interaction paradigms, with approximately 12 minutes allocated to each(based on the pilot study, to ensure all participants are familiar with the tool and have enough time to finish the visual creation task). Before each condition, participants watched a short demonstration video explaining the interaction paradigm and could ask clarification questions before beginning.
The order of the three paradigms was counterbalanced using a Balanced Latin Square to mitigate order effects. For each paradigm, participants were given four primary-level mathematics word problems covering addition, subtraction, multiplication, and division. Problems were sampled from a curated dataset of 173 textbook-derived examples.

After each condition, participants rated their experience using ten 7-point Likert items(Table~\ref{tab:tab1}) adapted from NASA-TLX\cite{hart1988development} and SUS\cite{brooke1996sus}. The items were reviewed and refined through internal pilot testing to ensure clarity and relevance to the study context. After completing all three conditions, they participated in a 10-minute semi-structured interview.

\subsection{Data Collection and Analysis}

We collected data from multiple sources, including pre-study screening questionnaires, post-task questionnaires, screen and audio recordings of study sessions, researcher observational notes, and system interaction logs (e.g., queries, tool actions, and generated images). 

After each condition, participants rated their experience using ten statements on 7-point Likert scales. Normality was assessed using Shapiro--Wilk tests, which indicated non-normal distributions for all items ($p < .05$). We therefore conducted Friedman tests to examine overall differences across the three conditions. When significant effects were observed, we performed post-hoc pairwise comparisons using Wilcoxon signed-rank tests with Bonferroni correction. Questionnaire results are reported in Table~\ref{tab:tab1}.

All study sessions were video-recorded, resulting in 1,609 minutes of recordings that captured participants’ interactions and think-aloud verbalizations. Recordings were transcribed and proofread prior to analysis. We conducted an inductive thematic analysis, beginning with in vivo coding to preserve participants’ own language\cite{saldana2021coding}, followed by iterative refinement into higher-level themes through discussions among the authors. Observational notes and interaction logs were analyzed alongside transcripts to capture patterns of tool use, user behavior, and breakdowns during visual creation. 


\section{Results}

\newcommand{\figurewidth}{8cm} 

\begin{table*}[t]
\centering
\caption{
Distribution of questionnaire responses across interaction paradigms (Pre-gen, Mid-gen, Post-gen).
M = mean; SD = standard deviation.
$p$-values are from Friedman tests (*$p<.05$; **$p<.01$).
Responses were collected on a 7-point Likert scale (1 = strongly disagree, 7 = strongly agree).
Q4 and Q5 (mental effort and frustration) were reverse-coded prior to statistical analysis and visualization, such that across all 10 items, higher scores consistently indicate more positive evaluations.
}

\label{tab:tab1}

\begin{tabular}{p{6cm} p{0.8cm} c c c p{6.6cm}}
\toprule
\textbf{Statement} & \textbf{Cond.} & \textbf{M} & \textbf{SD} & \textit{p} & \textbf{Agreement(dark red = 1, dark blue = 7)} \\
\midrule

\multirow{3}{6cm}{Q1: I found this tool easy to learn and use.}
& Pre  & 5.75 & 1.54 & \multirow{3}{*}{0.20}
& \multirow{3}{*}{\includegraphics[width=6.6cm]{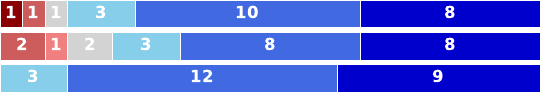}} \\
& Mid  & 5.58 & 1.56 & & \\
& Post & \textbf{6.25} & 0.68 & & \\
\midrule

\multirow{3}{6cm}{Q2: I felt confident creating visuals with this tool.}
& Pre  & 5.42 & 1.77 & \multirow{3}{*}{0.06}
& \multirow{3}{*}{\includegraphics[width=6.6cm]{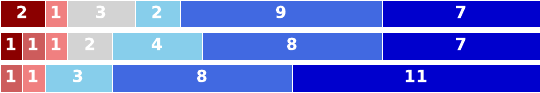}} \\
& Mid  & 5.46 & 1.64 & & \\
& Post & \textbf{6.04} & 1.30 & & \\
\midrule

\multirow{3}{6cm}{Q3: The tool behaved in ways I expected.}
& Pre  & 4.29 & 2.14 & \multirow{3}{*}{\textbf{0.00**}}
& \multirow{3}{*}{\includegraphics[width=6.6cm]{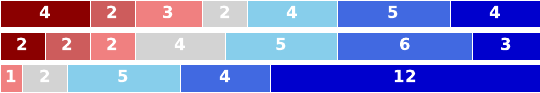}} \\
& Mid  & 4.58 & 1.82 & & \\
& Post & \textbf{6.00} & 1.22 & & \\
\midrule

\multirow{3}{6cm}{Q4: It required a lot of mental effort to use this tool effectively.}
& Pre  & 5.17 & 2.08 & \multirow{3}{*}{0.28}
& \multirow{3}{*}{\includegraphics[width=6.6cm]{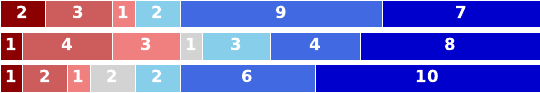}} \\
& Mid  & 4.88 & 2.09 & & \\
& Post & \textbf{5.50} & 1.87 & & \\
\midrule

\multirow{3}{6cm}{Q5: I felt frustrated while using this tool.}
& Pre  & 4.71 & 2.48 & \multirow{3}{*}{0.56}
& \multirow{3}{*}{\includegraphics[width=6.6cm]{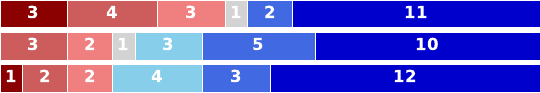}} \\
& Mid  & \textbf{5.46} & 1.82 & & \\
& Post & 5.54 & 1.93 & & \\
\midrule

\multirow{3}{6cm}{Q6: The image matched the number of objects for the math problem.}
& Pre  & 3.33 & 2.37 & \multirow{3}{*}{\textbf{0.00**}}
& \multirow{3}{*}{\includegraphics[width=6.6cm]{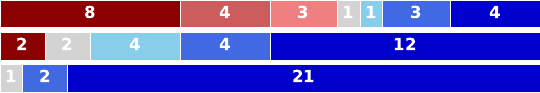}} \\
& Mid  & 5.75 & 1.78 & & \\
& Post & \textbf{6.79} & 0.66 & & \\
\midrule

\multirow{3}{6cm}{Q7: This tool helped me communicate the math concept effectively.}
& Pre  & 4.83 & 2.28 & \multirow{3}{*}{0.12}
& \multirow{3}{*}{\includegraphics[width=6.6cm]{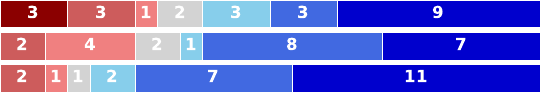}} \\
& Mid  & 5.25 & 1.75 & & \\
& Post & \textbf{5.83} & 1.58 & & \\
\midrule

\multirow{3}{6cm}{Q8: I feel comfortable using visuals from this tool in my teaching.}
& Pre  & 4.96 & 2.40 & \multirow{3}{*}{0.64}
& \multirow{3}{*}{\includegraphics[width=6.6cm]{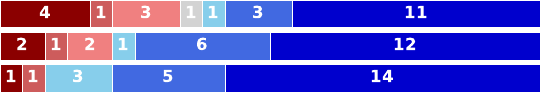}} \\
& Mid  & 5.62 & 2.02 & & \\
& Post & \textbf{6.08} & 1.59 & & \\
\midrule

\multirow{3}{6cm}{Q9: This tool gave me ideas for how to visualize the math scenario.}
& Pre  & \textbf{5.62} & 1.64 & \multirow{3}{*}{0.98}
& \multirow{3}{*}{\includegraphics[width=6.6cm]{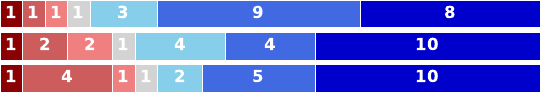}} \\
& Mid  & 5.38 & 1.91 & & \\
& Post & 5.25 & 2.09 & & \\
\midrule

\multirow{3}{6cm}{Q10: I would use a tool like this again for other teaching visuals.}
& Pre  & 5.29 & 2.10 & \multirow{3}{*}{0.93}
& \multirow{3}{*}{\includegraphics[width=6.6cm]{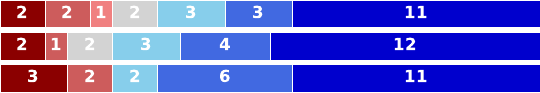}} \\
& Mid  & \textbf{5.62} & 1.93 & & \\
& Post & 5.42 & 2.21 & & \\
\bottomrule
\end{tabular}
\end{table*}

\subsection{Agency as a Prerequisite for Correctness-Sensitive Automation}

\subsubsection{Perceived numerical correctness differs by paradigm}
The majority of participants (18/24) emphasized that automation of visual generation is only valuable when it preserves their ability to verify and correct correctness-sensitive details. 
This tension was most evident in teachers’ concerns about numerical accuracy when generating instructional visuals. 
Quantitatively, participants' ratings for ``Q6: image matched the number of objects'' differed reliably across interaction paradigms ($Z = 24.83$, $p < .01$). Post-hoc Wilcoxon signed-rank tests revealed a clear ordering: post-gen was rated significantly higher than pre-gen ($Z = 0.0, p < .01$) and mid-gen ($Z = 3.5, p < .01$), indicating that teachers perceived post-generation control as most effective for preserving numerical correctness, despite its lower level of automation.

\subsubsection{Why post-gen felt safer: verifiability and localized correction}
One key reason for this preference was that post-gen control allowed teachers to retain agency over precise visual construction. This agency supported correctness in three concrete ways. First, it enabled explicit verification of visuals: 16 participants reported actively checking visual correctness while constructing or revising visuals. 
As P7 explained, \inlinequote{I prefer accuracy over speed --- I need to know this one is correct.} 
Besides, post-gen reduced the risk of unintended side effects during correction. 
Thirteen participants noted that local edits in post-gen (e.g., duplicating or deleting a single object) behaved predictably and did not trigger unexpected global changes. 
In contrast, seven participants  reported that pre-gen and mid-gen interactions often resulted in modifications to areas they did not intend to change. 
Finally, commonly used interactions reduced the effort required to maintain correctness. Seven participants noted that direct manipulation actions, such as drag-and-drop and copy–paste, enabled them to focus on visual accuracy rather than on managing the system, as these interactions align with those used in tools like PowerPoint and digital whiteboards.

In contrast, higher levels of automation were frequently described as fragile when they undermined correctness. Based on manual verification of the 90 images generated using pre-gen control across all participants, only 11 images contained the correct number of objects specified in the prompt (accuracy: 12.2\%). 
This issue was also noticed by participants: 17 participants reported that highly automated generation like pre-gen control paradigm could produce visually appealing but numerically incorrect images. 
Observational notes further showed that 11 participants abandoned otherwise promising images once inconsistencies were detected. As P6 summarized, \inlinequote{The quality is awesome --- I would absolutely love to use this, but only if the number of objects is correct.} Mid-gen control was often described as a partial improvement. Based on manual verification, 70 out of the 117 images generated using mid-gen control contained the correct number of objects (accuracy: 59.83\%). 
Fifteen participants appreciated that specifying layouts helped encode structure, yet still reported needing to verify the final image. In contrast, post-gen control exhibited substantially higher numerical accuracy: of the 93 images generated using post-gen control, only 2 were numerically incorrect (accuracy: 97.85\%). Together, these findings show that in correctness-sensitive tasks, automation is only effective when teachers retain sufficient agency to verify and correct generated visuals.

\subsection{Aligning System Behavior with Teacher Intent}
\subsubsection{Predictability ratings favor post-gen}
Teachers’ trust in an interaction paradigm depended largely on how closely system behavior aligned with their expectations, particularly in how user actions were reflected in the resulting visuals. 
This was reflected in ratings for ``Q3: The tool behaved in ways I expected'', which differed reliably across interaction paradigms ($Z = 19.98, p < .01$). 
Post-gen control was rated significantly higher than both pre-gen ($Z = 11.0, p < .01$) and mid-gen control ($Z = 12.0, p < .01$), while no statistically significant difference was observed between pre-gen and mid-gen control. 

\subsubsection{Why post-gen felt predictable, and why pre-/mid-gen felt uncertain}
Across sessions, teachers repeatedly emphasized the need for a clear and inspectable mapping between their input and the system's output.
Sixteen participants explicitly stated that predictable behavior was critical for efficiently creating visuals, especially in mathematics education where \inlinequote{small deviations can undermine pedagogical intent}, as P6 suggested. Post-gen control supported this claim by allowing teachers to directly observe how each action affected the visual. 
Twelve participants noted that operations such as moving, duplicating, or deleting objects produced immediate and localized changes, making system behavior easy to anticipate. As P15 explained, \inlinequote{I can create everything even from scratch and not be influenced by the AI’s error, so there is no additional cost of modification.} 
This directness reduced uncertainty during iteration: 16 participants reported feeling confident making incremental adjustments without worrying that automated processes would introduce new errors elsewhere in the visual.

By comparison, uncertainty in pre-gen and mid-gen control arose from difficulty anticipating how the system would interpret user intent. With pre-gen control, 12 participants reported that although they could often identify what was wrong in a generated image, they were unsure how to phrase corrections or whether a revised prompt would resolve the issue without creating new problems. P13 and P17 described this experience as \inlinequote{guessing how the system will respond} rather than editing. 
In mid-gen control, uncertainty shifted to the realization of layouts during generation. 
Seven participants described cases in which elements were placed in unexpected regions or objects were missing, requiring repeated attempts after the initial generation. As P9 noted, \inlinequote{Even if the layout looks right, I still have to check what the system actually produces.}

Overall, these findings indicate that a transparent intent–outcome relationship was central to how teachers evaluated system behavior, helping explain why post-gen control received higher ratings on expected behavior despite otherwise similar usability ratings.

\subsection{Tool Preference Is Stage-Dependent and Motivates Hybrid Workflows}
\subsubsection{No single paradigm dominates overall}
Rather than converging on a single optimal interaction paradigm, teachers' evaluations reflected stage-dependent preferences across the visual creation workflow. Quantitatively, Friedman tests revealed no significant differences between tools on eight of the ten questionnaire items, including measures related to ease of use, mental effort, educational fit, and reuse intention. 
This absence of significant differences suggests that no single paradigm consistently outperformed the others across all subjective dimensions. Instead, the tools appeared to offer different strengths that were valued in different situations, which helps explain why interview responses reflected context-dependent preferences rather than a single dominant choice.

\subsubsection{Mapping Interaction Paradigms to Authoring Stages}
Our findings further show that teachers associated specific interaction paradigms with distinct stages of instructional workflow. 
Pre-gen control was most often described as useful during early ideation, when teachers sought inspiration or visually engaging representations. 
Correspondingly, pre-gen received the highest mean score on ``Q9: This tool gave me ideas for how to visualize the math scenario''. 
Seven participants emphasized its value for exploration. 
As P6 noted, \inlinequote{It's a nice teammate for brainstorming, and I like the idea that it tries to scaffold the question to help students understand.} Similarly, P8 described using it when they \inlinequote{have time to explore ideas or make posters for motivating students.} 
Mid-gen was commonly associated with structuring representations, especially for tasks involving spatial organization. 
Seven teachers described using layouts to scaffold multiplication problems before refining the visual. 
Post-gen, by contrast, was consistently framed as the safest option for final classroom materials. 
Four participants emphasized its suitability for \inlinequote{quickly preparing worksheet exercises}.

\subsubsection{Stage-Specific Strengths Motivate Hybrid Workflows}
This stage-dependent reasoning also motivated interest in hybrid workflows that combine multiple interaction styles across authoring stages. Rather than committing to a single paradigm, 15 teachers explicitly requested systems that support automated generation from prompts followed by manual, object-level refinement. 
For example, P13 described an ideal tool as one where you \inlinequote{type what you want in text, like the pre-gen control tool, then have multiple alternative images, select your preferred one, and edit directly like the post-gen control tool.} 
Others emphasized combining conversational input with direct manipulation. P4 noted that it was \inlinequote{hard to decide a single tool that dominates}, expressing a preference for combining conversational generation with drag-and-drop editing. 
P20 further described an ideal system that integrates \inlinequote{a bank of images, region-level editing, and object-level editing for dragging and resizing}. 
Together, these findings illustrate how teachers' preferences were shaped less by overall tool superiority and more by how well different interaction styles could be composed to support distinct stages and use cases within the visual creation process.

\section{Discussion}

\label{sec:discussion}
In correctness-sensitive educational settings, the timing of teacher intervention critically shapes AI-assisted visual generation.
Building on prior work on mixed-initiative interaction~\cite{horvitz1999principles,allen1999mixed,horvitz2004busybody,lehmann2023mixed}, we interpret this insight through the lens of interaction timing, arguing that pre-generation, mid-generation, and post-generation control give rise to qualitatively different workflows, error-handling strategies, and trade-offs between automation and human agency.
These differences are particularly consequential in primary mathematics education, where visual representations must reliably support instructional intent and classroom practice~\cite{eilam2012teaching,clark2010graphics}.


\subsection{Control Timing Reframes the Nature of Risk in Generative Authoring}

A central insight from our findings is that different interaction conditions do not simply change how much effort teachers expend; they fundamentally reshape where risk resides in the authoring process. Across conditions, risk shifts from interpreting pedagogical intent, to translating intermediate structure, to managing local manual refinement. 
This shift reflects how the timing of human control determines whether errors are opaque, partially inspectable, or fully localizable during AI-assisted visual authoring.

In \emph{pre-generation control}, risk is concentrated in model interpretation. Conversational prompting enables rapid initiation, but correctness depends on the system's ability to faithfully interpret linguistic intent and constraints\cite{zamfirescu2023johnny,liu2022design}. As a result, teachers frequently entered cycles of generating, inspecting, and re-prompting when outputs appeared visually plausible yet contained errors. 
Verification effort thus became unavoidable, and repeated breakdowns in intent–output alignment undermined trust in the generated visuals.
This interaction pattern can be understood through Norman's gulfs of execution and evaluation~\cite{norman2013design}, which characterize the distance between users' goals and available system actions, as well as the difficulty of interpreting system state relative to those goals. 
In pre-generation workflows, teachers must translate pedagogical intent into natural-language prompts, widening the gulf of execution, while the lack of transparency in how prompts are translated into images makes it difficult for teachers to understand why outputs diverge from their intended representation.

\emph{Mid-generation control} relocates part of this risk into an inspectable intermediate representation. In line with distributed cognition theory~\cite{hollan2000distributed}, editable layouts function as external representations that reduce internal cognitive burden and support inspection of spatial and grouping relationships. By externalizing structure, teachers can express pedagogical intent more explicitly, reducing ambiguity during specification. However, this condition introduces a different form of uncertainty: whether the approved layout will be faithfully realized in the final rendered image. Teachers frequently described the need to re-verify outputs despite having validated the layout, suggesting that risk shifts from interpreting intent to translating between intermediate and final representations.

\emph{Post-generation control} minimizes interpretation and translation risk by making system state continuously visible and directly editable. Rather than negotiating intent with the model, teachers can verify and correct errors locally, keeping the cost of correction bounded and predictable. This localization of error reduces both cognitive and emotional cost during visual authoring~\cite{hollan2000distributed}. Importantly, this advantage does not stem merely from providing ``more control,'' but from eliminating hidden transformations after generation. Teachers can directly observe how each action affects the visual outcome, enabling immediate evaluation and correction.

Several participants further noted that post-generation workflows increased their sense of accountability: manual interaction encouraged them to actively take responsibility for the final representation. This suggests that control timing reshapes perceived responsibility in addition to effort and accuracy~\cite{shneiderman2022human}. Consistent with prior findings that even limited opportunities for user intervention can reduce algorithm aversion~\cite{dietvorst2018overcoming,logg2019algorithm}, post-generation control allowed teachers to regain trust by making correction local, transparent, and reliable.

Taken together, these patterns suggest that control timing reshapes not only effort, but also how teachers encounter, interpret, and manage errors in AI-assisted visual authoring workflows.

\subsection{From Visual Quality to Auditability in Correctness-Sensitive Domains}

Across interaction conditions, teachers evaluated AI-generated visuals through a lens that differs markedly from many creative or exploratory domains. In primary mathematics instruction, visual representations are not decorative artifacts; they operationalize quantities, relations, and structures that students are expected to reason about. 
Prior learning science research shows that such representations actively shape conceptual understanding, particularly in mathematics where meaning is distributed across symbolic and visual forms~\cite{duval1999representation,pape2001role,presmeg2006research}. 
Consequently, teachers treated even small numerical inconsistencies as unacceptable, as incorrect or misleading visuals risk introducing or reinforcing misconceptions that can persist even after verbal correction~\cite{rau2017conditions}. 
Rather than judging outputs by visual appeal or prompt alignment alone, teachers implicitly evaluated generated visuals based on their representational fidelity: whether the visual structure preserved the intended mathematical relationships, quantities, and groupings~\cite{duval1999representation,rivera2011toward}. 
Within this framing, post-generation control was valued in part because it supported auditability: the ability to directly inspect, verify, and enforce correctness without needing to infer hidden system behavior.
By contrast, pre-generation and mid-generation workflows often required teachers to double-check outputs or speculate about how to phrase corrections, an effort many participants viewed as undermining the promise of AI assistance.

Prior work suggests that such inspectability is not merely a preference but a core usability requirement for human–AI systems, particularly in high-stakes contexts~\cite{amershi2019guidelines}. When users cannot easily verify system outputs, they are forced to engage in compensatory strategies, such as repeated prompting, that impose extraneous cognitive load unrelated to the primary task~\cite{mayer2005cognitive}. 
In our study, teachers explicitly described prompt negotiation and repeated verification as mentally demanding and pedagogically distracting.

These demands for auditability persist even as text-to-image models continue to improve. Although recent generative models demonstrate strong performance in artistic and creative domains~\cite{ramesh2021zero,rombach2022high}, prior work has shown that they remain prone to systematic errors in tasks requiring precise numerical reasoning or structural fidelity~\cite{kajic2024evaluating,cao2025text}. Our findings suggest that providing richer inputs, such as detailed prompts, problem statements, or intermediate layout specifications, may reduce ambiguity but does not eliminate discrepancies between teachers’ pedagogical intent and generated outputs. Rather than viewing these discrepancies solely as model shortcomings, they highlight the importance of interaction designs that allow users to inspect, verify, and correct outputs when model behavior diverges from instructional goals.

Taken together, these findings indicate that in correctness-sensitive domains such as mathematics education, output quality alone is an insufficient success criterion. Auditability emerges as a foundational design requirement.
By supporting verifiability, systems enable teachers to appropriately calibrate trust, maintaining productive reliance on AI assistance while retaining responsibility for correctness.

\subsection{Why No Single Condition Wins: Stage-Dependent Needs in Teacher Workflows}

Rather than favoring a single interaction condition, teachers' preferences varied across different stages of the instructional workflow.
Our findings from the user study suggest that teachers tended to associate different interaction conditions with different stages of the authoring process.
Pre-generation control was primarily valued during early ideation, when teachers were exploring possible visual formulations and visual variety supported rapid exploration rather than precision.
Mid-generation control was valued during the structuring stage, where an intermediate representation made grouping and spatial relations explicit before final rendering, supporting inspection and revision~\cite{russell1993cost,pirolli2005sensemaking}.
Post-generation control was most strongly preferred during finalization, when teachers prioritized correctness and predictability and sought bounded, local corrections rather than iterative prompt negotiation.

This stage-dependent pattern can be understood as an exploration--exploitation trade-off~\cite{march1991exploration}. Pre-generation control facilitated exploration by enabling the generation of diverse candidates, while post-generation control supported exploitation by allowing reliable refinement of a chosen representation. As authoring progressed and the pedagogical consequences of errors increased, teachers increasingly favored interaction modes that made system behavior predictable and errors explicit.

Taken together, these findings suggest that no single interaction condition can optimally support all stages of instructional authoring. Designing for such workflows enables systems to align with teachers' evolving goals, from exploration and sensemaking to verification and classroom-ready production.

\subsection{Design Takeaways for Correctness-Sensitive Generative Tools}
We derive design takeaways for generative authoring in correctness-sensitive contexts. While prior human--AI guidelines emphasize transparency, correction, and trust \cite{amershi2019guidelines}, our findings show that in visual authoring these depend on \textbf{when control is exercised}, as timing shapes how uncertainty arises and how errors can be inspected and corrected.

    \subsubsection{Make correctness explicit}
    In correctness-sensitive educational contexts, generative tools should make correctness visible and open to inspection, rather than treating it as a hidden outcome of generation. When key aspects of a visual's structure and meaning are legible to teachers, they can more easily assess whether the output aligns with instructional intent and classroom requirements.

    \subsubsection{Support precise, localized correction}
    When teachers identify problems in generated outputs, systems should allow them to address those issues directly without causing unintended changes elsewhere. Designing for bounded, local correction helps preserve teachers' work, reduces unnecessary iteration, and supports confident refinement of instructional materials.

    \subsubsection{Expose structure without introducing unnecessary overhead}
    Structural representations can play an important role in supporting inspection and revision, but only when they remain lightweight and predictable. Generative tools should help users reason about structure without requiring extensive configuration or introducing new sources of uncertainty.

    \subsubsection{Design for stage-aware authoring workflows}
    Teachers' needs evolve across different stages of instructional authoring. Rather than optimizing for a single interaction paradigm, generative tools should support multiple forms of control that align with exploration, structuring, and finalization, allowing users to move fluidly between stages as their goals change.

\subsection{Limitations and Future Work}

This study focuses on single-session use in a controlled setting and therefore primarily captures teachers' first-use experiences rather than longer-term adoption into authentic lesson-planning workflows. As a result, we cannot assess how teachers' strategies or verification practices evolve over time.
Future work should examine longitudinal and in-situ deployments to understand how these tools fit within realistic planning routines and institutional contexts.

Our tasks center on generating visuals for primary-level arithmetic, which constrains the generalizability of our findings. Interaction preferences and correctness sensitivities may differ for other grade levels or subjects. 
Comparative studies across domains could clarify when automation, structured representations, or direct manipulation are most beneficial, and inform systems that adapt interaction support to task-specific pedagogical goals.

While we compare teachers' perceptions and observed workflows across interaction conditions, we do not evaluate downstream instructional or learner outcomes. Consequently, we cannot determine whether the observed differences translate into higher-quality instructional materials or improved student understanding. 
Future work should connect authoring workflows to classroom use, examining how different interaction conditions affect material accuracy, instructional clarity, and student learning.

Finally, although teachers repeatedly emphasized the importance of correctness, verification in our study largely relied on manual inspection. This places sustained cognitive demands on teachers, particularly in domains like mathematics. Future systems should therefore support verifiable correctness explicitly through constraint-aware generation, automatic consistency checks, and inspectable ``sanity check'' views.
Interactive debugging mechanisms that localize errors and suggest targeted fixes could further reduce verification effort and increase instructional confidence.

\section{Conclusion}
This paper investigates how the timing of human control shapes teachers' use of generative AI for creating educational visuals in correctness-sensitive settings. 
Through a mixed-methods study with primary mathematics teachers, we show that different points of intervention in the generation process lead to qualitatively different authoring workflows and ways of managing error. 
Our findings indicate that in domains where visual representations carry instructional meaning, the value of generative systems depends not only on the quality or efficiency of their outputs, but on whether interaction designs make verification and responsibility manageable in practice. 
Rather than converging on a single optimal interaction condition, teachers' preferences reflect stage-dependent needs that evolve from exploration to refinement and finalization. These results suggest that future educational authoring tools should support hybrid, stage-aware workflows that integrate automation with transparent, inspectable forms of control, enabling educators to reliably align AI-generated visuals with pedagogical intent and classroom use.

\begin{acks}
This research was supported by the ETH AI Center through a doctoral fellowship to Junling Wang, the Swiss AI Large Grant SCR1089274, and Swiss AI Small Grant \#63.
Additionally, the authors thank the reviewers, members of the PEACH Lab at ETH Zurich, and the participants in the study.
\end{acks}

\bibliographystyle{ACM-Reference-Format}
\balance
\bibliography{references}

\end{document}